\documentstyle[onecolumn,epsfig]{mn}
\newcommand{\etal}{{et~al.~}}
\newcommand{\uk}{u_k}
\newcommand{\bi}[1]{\mbox{\boldmath $#1$}}
\newcommand{\bay}[2]{\left(\begin{array}{c}\mbox{$#1$} \\ \mbox{$#2$}
\end{array}\right)} 
\title{Spawning and merging of Fourier modes and phase coupling in 
cosmological density bispectrum} 
\author[Lung-Yih Chiang]{Lung-Yih Chiang\thanks{E-mail : {\tt chiang@tac.dk}}\\
Theoretical Astrophysics Center, Juliane Maries Vej 30,
DK-2100,  Copenhagen, Denmark}

\date{Accepted 2003 ???? ???; Received 2003 ???? ???}

\begin{document}
\maketitle

\begin{abstract}
In the standard picture of cosmological structure formation, initially
random-phase fluctuations are amplified by non-linear gravitational
instability to produce a final distribution of mass which is highly
non-Gaussian and has highly-coupled Fourier phases. We use the
Zel'dovich approximation in one dimension to elucidate the onset of
non-linearity including mode spawning, merging and coupling. 
We show that as gravitational clustering proceeds, Fourier modes are
spawned from parent ones, with their phases following a harmonic
relationship with the wavenumbers. Spawned modes could also
merge leading to modulation of the amplitudes and phases which
consequently breaks such harmonic relation. We also use simple toy models to
demonstrate that bispectrum, Fourier transform of connected
three-point correlation functions, measures phase coupling at most at
the second-order only when the special wavenumber-phase harmonic
relation holds. Phase information is therefore partly registered in
bispectrum and it takes a complete hierarchy of polyspectra to fully
characterize gravitational non-linearity.

\end{abstract}


\section{Introduction}
One of the main issues in cosmology is quantitative
characterization of the large-scale structure of the
Universe. The Universe from present observations is highly
non-linear on scales up to roughly 8$h^{-1}$Mpc. In the framework of
the inflation paradigm, the structure we see today is blown up from tiny
inhomogeneities originated from quantum fluctuations. The cosmic
microwave background has provided a strong evidence that the Universe
was fairly homogeneous and isotropic with fluctuations one part in
$10^5$ in the past. Such initial density fluctuations in the simplest
inflationary universe constitute a Gaussian random field. 

Gaussian random Fields \cite{bbks} are useful because of its analytical
simplicity. One particularly interesting property of Gaussian random
fields is that the real and imaginary part of the Fourier modes are
both Gaussian distributed and mutually independent, or in other
words, the Fourier phases are randomly distributed between 0 and $2
\pi$. The statistical properties are then completely specified by its
second-order statistics: its two-point correlation function $\xi(r)$
(with zero connected $n$-point correlation functions for $n>2$), or
alternatively, its power spectrum $P(k)$. 

In the framework of gravitational instability, a perturbative method
can be adopted at the early stage of density clustering (see
e.g. Peebles 1980, Bernardeau et al. 2002). The linear perturbation
theory is applicable when the density fluctuations are small i.e. its
variance of density contrast $\langle \delta^2 \rangle <1$. The
statistical distribution of the density field such as originally
random phases remains invariant in the linear regime, except its
variance increasing with time. 

The departure of the density field from the linear regime gives
rise to phase coupling. Second-order statistics, such as power
spectrum and two-point correlation function, throw
away the fine details of the delicate pattern of cosmic
structure. These details lie in the distribution of Fourier phases to
which second-order statistics are blind. The evident shortcomings of
$P(k)$ can be partly ameliorated by defining higher-order quantities
such as the bispectrum~\cite{peebles,goroff,bernardeau,hivon,mvh,scoccimarro,setal,scf,vwhk}
or correlations of $\delta({\bi k})^2$ \cite{stir}. Higher-order
correlations and polyspectra find this information term by term so
that an infinite hierarchy is required for a complete statistical
characterization of the fluctuation field.

Due to the close connection between morphology and Fourier phases
\cite{c3}, there have been attempts in investigating the behaviour of phases
\cite{ryden,sodasuto,jain}. They focus on phase shift from the
original one. Scherrer, Melott and Shandarin \shortcite{sms} develop
a practical method on quantifying phase coupling.
Chiang \& Coles \shortcite{phaseentropy} take phase difference between
neighbouring modes on Shannon entropy to quantify phase information. Due to the
circular nature of phases, a novel visualization method is also
developed \cite{visualization}. Chiang, Coles \& Naselsky
\shortcite{mapping} and Chiang, Naselsky \& Coles
\shortcite{chisquare} develop a novel method, return
mapping of phases to render phases onto a `return map' to quantify phase
associations between different $\Delta {\bi k}$. Watts, Coles \&
Melott \shortcite{wcm} have shown that the probability distribution of
phase difference between neighbouring modes displays a universal
behaviour for clustering phenomenon.  

Since both Fourier phase coupling and non-zero higher-order
correlation functions depict gravitational non-linearity, Watts \&
Coles \shortcite{wc}; Matsubara \cite{matsubara} have shown the
relationships between phase coupling and bispectrum which displays the
lowest order non-linearity in density perturbations. Hikage, Matsubara
\& Suto \shortcite{hikage} have used phase sum from wavenumbers (${\bi
k}$ vectors) forming a triangle as a descriptor of non-linear
gravitational clustering.

Despite the progress described above, efforts are yet to be made in
understanding the direct relationship between gravitational clustering and
phase coupling. In this paper we use one-dimensional
Zel'dovich approximation as a clustering scheme to investigate mode
coupling when gravitational perturbations depart from the linear regime. This
moment is related to the so-called {\it quadratic density fields}
(Coles \& Barrow 1987; Watts \& Coles 2002 and references therein), in which the quadratic term induces non-linearity. The
non-linearity caused by gravitational clustering can be described in
two effects: {\it mode spawning} by the quadratic and higher-order terms and
{\it mode merging} leading to modulation of the amplitudes and
phases. These two effects are in action as cosmological density
clustering proceeds.

This paper is organized as follows. In Section 2 we re-visit some
useful technical background: the
commonly-used statistical tools and theories: the power spectrum and two-point
covariance functions, the linear theory of cosmological density
perturbations, and the Zel'dovich approximation which in one dimension
provides the insight on phase coupling. In Section 3 we
use such one-dimensional model as a clustering scheme for analyses of
the onset of the following phenomena: mode spawning and merging from
gravitational clustering and their relationships with bispectrum. We
further examine these phenomena by direct 1D simulations into
highly non-linear regime in Section 4. The discussions are in Section
5. For completeness in the Appendix we prove the exactness of the density
evolution from the 1D Zel'dovich approximation.  

\section{Technical background}
\subsection{Power spectrum and two-point covariance functions}
The mathematical description of an inhomogeneous Universe revolves
around the dimensionless density contrast, $\delta({\bi x})$,
which is obtained from the spatially-varying matter density
$\rho({\bi x})$ via
\begin{equation} \delta ({\bi x}) =
\frac{\rho({\bi x})-\rho_0}{\rho_0},
\end{equation}
where ${\bi x}$ is the comoving coordinate, $\rho_0$ is the global
mean density \cite{peebles}. The two-point covariance function, which
measures the excess probability over Poisson distribution between a
pair with separation ${\bi r}$, is defined as
\begin{equation}
\xi({\bi r})=\langle \delta({\bi x}) \delta({\bi x} +{\bi r}) \rangle,
\end{equation}
where the mean is taken over all points ${\bi x}$. It is also useful to
expand the density contrast in Fourier series, in which $\delta$ is
treated as a superposition of plane waves: 
\begin{equation}
\delta ({\bi x}) = \sum \delta({\bi k}) \exp(\imath {\bi k}\cdot
{\bi x}). \label{eq:fourier}
\end{equation}
The Fourier transform $\delta({\bi k})$ is complex and
therefore possesses both amplitude $|\delta ({\bi k})|$
and phase $\phi_{\bi k}$ where
\begin{equation}
\delta({\bi k})=|\delta ({\bi k})| \exp(\imath \phi_{\bi k}).
\label{eq:fourierex}
\end{equation}
The power spectrum is defined as
\begin{equation}
\langle \delta({\bi k}_1) \delta({\bi k}_2) \rangle = (2 \pi)^3 P(k)
\delta^D({\bi k}_1+{\bi k}_2), \label{eq:powersp}
\end{equation}
which is the Fourier transform of the two-point covariance function
via Wiener-Khintchin theorem. We can analogously define the bispectrum
as the third-order moment in Fourier space:
\begin{equation}
\langle \delta({\bi k}_1) \delta({\bi k}_2) \delta({\bi k}_3) \rangle =
(2 \pi)^3 B({\bi k}_1,{\bi k}_2,{\bi k}_3) \delta^D({\bi k}_1+{\bi
k}_2+{\bi k}_3 ), \label{eq:bisp}
\end{equation}
which is the Fourier transform of the connected three-point covariance
function. 

\subsection{The linear theory of density perturbations}
Gravitational instabilities is believed to be the driving force of
large-scale structure formation. When the density
perturbation is small, evolution of the density contrast can be
obtained analytically through the {\it linear perturbation theory} from 3
coupled partial differential equations. They are the linearized continuity
equation,
\begin{equation}
{\partial\delta\over \partial t} = - {1\over a}{\bi \nabla_x}\cdot
{\bi v}, \label{eq:lCont}
\end{equation}
the linearized Euler equation
\begin{equation}
{\partial {\bi v}\over\partial t} + {\dot a\over a}{\bi v} = -
{1\over \rho a}{\bi \nabla_x} p -{1\over a}{\bi \nabla_x}\phi,
\label{eq:lEuler}
\end{equation}
and the linearized Poisson equation
\begin{equation}
{\bi \nabla_x}^2\phi = 4\pi G a^2\rho_0\delta. \label{eq:lPoisson}
\end{equation}

In these equations, $a$ is the expansion factor, $p$ is the pressure,
${\bi \nabla_x}$ denotes a derivative with respect to the comoving
coordinates ${\bi x}$, ${\bi v}=a \dot{\bi x}$ is the peculiar
velocity and  $\phi({\bi x},t)$ is the peculiar gravitational
potential. From Eq.(\ref{eq:lCont})-(\ref{eq:lPoisson}), and if one
ignores pressure forces, it is easy to obtain an equation for the
evolution of $\delta$:
\begin{equation}
\ddot\delta + 2(\frac{\dot a}{a})\delta - 4 \pi G \rho_0 \delta = 0.
\label{eq:2ndorder}
\end{equation}
For a spatially flat universe dominated by pressureless matter,
$\rho_0(t) = 1/6\pi Gt^2$ and Eq.(\ref{eq:2ndorder}) admits two
linearly independent power-law solutions 
\begin{equation}
\delta({\bi x},t) = b_{\pm}(t)\delta_0({\bi x}), \label{eq:linearsol}
\end{equation} 
where $\delta_0({\bi x})$ is the initial density distribution, $b_+(t) \propto
a(t) \propto t^{2/3}$ is the growing mode and $b_{-}(t) \propto
t^{-1}$ is the decaying mode.

As one can see from Eq.(\ref{eq:linearsol}), the growth depends only
on time, so the growth of density distribution in the linear regime
does not alter the phases.

\subsection{The Zel'dovich Approximation}
In order to examine analytically non-linear effects induced by gravitational
clustering, we use the Zel'dovich approximation (1970) as a clustering
scheme, which extrapolates the evolution of density
perturbations into non-linear regime. This extrapolation from the
linear theory follows the perturbation in particle trajectories
rather than in density fields. In the Zel'dovich approximation (ZA), a
particle initially placed at the Lagrangian coordinate ${\bi q}$ is
perturbed after a time $t$ to an Eulerian coordinate ${\bi x}$. The
displacement of the particle simply depends on the constant velocity the
particle has when it is kicked off its initial position, and can be
written as $b(t){\bi u}({\bi q})$, so that
\begin{equation}
{\bi r}({\bi q},t)=a(t){\bi x}({\bi q},t)=a(t)[{\bi q}+b(t){\bi u}({\bi q})],
\end{equation}
where ${\bi r}$ is the resultant physical coordinate, $a(t)$ is the expansion factor and $b(t)$ is the growing
mode $b_{+}(t)$ from the linear perturbation theory. According to this
prescription, each particle moves with a constant velocity
along a ballistic trajectory, which resembles Newtonian inertial
motion. Note that the peculiar velocity
according to the ZA is $a\dot{\bi x}=a(t) \dot{b}(t) {\bi u}({\bi
q})$. For an irrotational flow, ${\bi u}({\bi q})$ can be expressed as
a gradient of some velocity potential $-\nabla \Phi_{0}({\bi q})$. 

We focus particularly on the applications of one-dimensional ZA. 
The ZA in 1D provides an exact solution of
density evolution \cite{buchert} in that the evolution of planar
collapse from the ZA has the same solution as from the Poisson equation 
until shell-crossing \cite{padmanabhan}. The ZA in 1D is simplified to 
\begin{equation}
x(q,t)=q+b(t) u(q) = q-b(t)\frac{d\Phi_0(q)}{dq}. \label{eq:1d}
\end{equation}
The density contrast can be derived from the conservation of mass $\rho dx=\rho_0 dq$  :
\begin{eqnarray}
\delta & = & \frac{\rho}{\rho_0}-1=\left(\frac{\partial x}{\partial
q}\right)^{-1}-1  = \left[1-b(t)\frac{d^2\Phi_0(q)}{dq^2}
\right]^{-1}-1 \label{eq:delta1d} \\
& = & \sum_n b^n(t)\left( \frac{d^2
\Phi_0}{dq^2}\right)^n. \label{eq:expansion}
\end{eqnarray}
The velocity potential $\Phi_0(q)$ can always be disintegrated as
\begin{equation}
\Phi_0(q)=\sum_i A_i \cos(\lambda_i q + \alpha_i).\label{eq:potential}
\end{equation}
One can recover the solution to the linear theory from Eq.(\ref{eq:delta1d}) by taking
partial differentiation $\partial/\partial t$ on both sides. Substituting $d^2 \Phi_{0}/dq^2$ with $b^{-1}\delta/(\delta+1)$, we reach
\begin{equation}
\frac{\partial\delta}{\partial t}= \frac {\dot
b}{b}(\delta+\delta^{2}). \label{eq:recover}
\end{equation}

Equation (\ref{eq:recover}) provides us the insight into phase coupling from
gravitational clustering. One important property of
Eq.(\ref{eq:recover}) is that the $\delta^{2}$ term is the ``culprit''
of the onset of non-linearity. Without the $\delta^2$ term we obtain
the same solution as that from the linear 
theory, $\delta\propto b$ ($\propto t^{2/3}$ if the Universe is
matter-dominated), i.e. density fields grow only with time in the
linear regime where the phases are unchanged (and uncorrelated if
primordial Gaussianity is assumed). It is
this quadratic term $\delta^{2}$ that the density field breaks away
from the linear regime and produces gravitational non-linearity, hence
phase coupling. In Eq.(\ref{eq:recover}), the quadratic term
$\delta^2$ is related to the ``quadratic density field''
\cite{cb,wc}.

\section{Mode Coupling and bispectrum}
In this section we use 1D ZA to demonstrate spawning and
merging of Fourier modes and phase coupling induced by gravitational clustering and its
relation with bispectrum. For the analyses in this Section, we
consider only an early stage of evolution, i.e. $b(t) \ll 1$ and
assume that the curvature of the velocity potential is small such that
$x \simeq q$ and $dx \simeq dq$. Fourier transform can then be
performed in the Lagrangian coordinate $q$. Direct simulations of 1D
ZA for gravitational evolution is in Section~\ref{simulation}.

\subsection{Mode spawning and mode merging}
Upon using 1D ZA, we firstly choose as a toy model the velocity potential \footnote{We put
minus sign for both cosine functions in the velocity potential to make the analytical form neat,
assuming both $A_1$ and $A_2$ positive, otherwise the phases would
have a shift by $\pi$.}
\begin{equation}
\Phi_0(q) = - A_1 \cos(\lambda_1 q + \alpha_1)- A_2 \cos(\lambda_2 q +
\alpha_2).
\end{equation}
When $b(t) \ll 1$, the density contrast can be expressed in terms of
only the first order from  Eq.(\ref{eq:expansion})
\begin{equation}
\delta \;\;\; \simeq \;\;\;\delta^{(1)} =  b(t)\left( \frac{d^2
\Phi_0}{dq^2}\right)= a_1 \cos(\lambda_1 q + \alpha_1) + a_2 \cos( \lambda_2 q+ \alpha_2)\equiv a_1 \bay{\lambda_1}{\alpha_1} + a_2
\bay{\lambda_2}{\alpha_2},
\end{equation}
where the round brackets hereafter denote cosine functions, $a_1= b(t) A_1
\lambda^2_1$ and $a_2= b(t) A_2 \lambda^2_2$. After Fourier transform
in Lagrangian coordinate we have only two modes with wavenumbers
$\lambda_1$ and $\lambda_2$, and phases $\alpha_1$ and $\alpha_2$,
respectively. So these are the only Fourier modes the moment
the clustering process begins (which we call the ``parent modes''). It is clear that the first order does not
display phase coupling. When the second-order term becomes comparable,   
\begin{eqnarray}
\delta & \simeq & \delta^{(1)} + \delta^{(2)} = b(t)\left( \frac{d^2 \Phi_0}{dq^2}\right) + b^2(t)\left( \frac{d^2 \Phi_0}{dq^2}\right)^2 \nonumber \\
       & = & \frac{a_1^2+a_2^2}{2} + a_1 \bay{\lambda_1}{\alpha_1} +
       a_2 \bay{\lambda_2}{\alpha_2} 
       + a_1^2 \bay{2 \lambda_1}{2
       \alpha_1}+ a_2^2 \bay{2
       \lambda_2}{2 \alpha_2} + a_1 a_2
       \bay{\lambda_1+\lambda_2}{\alpha_1+\alpha_2}+ a_1 a_2
       \bay{\lambda_1-\lambda_2}{ \alpha_1-\alpha_2}. \label{eq:2order}
\end{eqnarray}
Since the density contrast in 1D ZA can be expressed as a power series
of $d^2 \Phi_0/dq^2$ in Eq.~(\ref{eq:expansion}), the second-order term reflects the quadratic
density fields \cite{cb,wc}. This second-order term spawns modes with
wavenumbers that are from combinations of any 2 wavenumbers from the
parent ones, i.e. Fourier modes with wavenumbers $2\lambda_1$, $2\lambda_2$, $\lambda_1+\lambda_2$ and
$\lambda_1-\lambda_2$ spawned from combination of $\lambda_1$ and
$\lambda_2$. One important feature of quadratic density fields is that the
phases of the spawned modes follow the same kind of harmonic
relationship as the spawned wavenumbers, which we call hereafter
``wavenumber-phase harmonic relation''. Such relation subsequently forms phase
associations between Fourier modes and is crucial for bispectrum analysis. 

We can generalize the velocity potential in 1D ZA as a sum of cosine functions
$\Phi_0= - \sum_i A_i \cos( \lambda_i q + \alpha_i)$. The density
contrast of the first order and that up to the second order now become, respectively, 
\begin{equation}
\delta^{(1)} = \sum_i a_i \cos(\lambda_i+ \alpha_i) \equiv \sum_i a_i \bay{\lambda_i}{\alpha_i},
\end{equation}
where $a_i= b(t) A_i \lambda_i^2$, and
\begin{equation} 
\delta^{(1)} + \delta^{(2)}= \sum_i a_i \bay{\lambda_i}{\alpha_i} + \sum_{jk}
\frac{a_j a_k}{2} \left[\bay{\lambda_j - \lambda_k}{\alpha_j -
\alpha_k} + \bay{\lambda_j + \lambda_k}{\alpha_j + \alpha_k}
\right]. \label{eq:quadraticfield}
\end{equation}
If primordial Gaussianity is assumed, the phases
$\alpha_i$ of the initial velocity potential are uniformly random and
independently distributed between 0 and $2\pi$. Below we
categorize the effects induced by quadratic density fields.

\begin{itemize}
\item {\bf Mode spawning :} In Eq.(\ref{eq:quadraticfield}) beyond the
first order, new modes are spawned from the $\delta^{(2)}$ and
correlated phases are created following the spawned
modes. The wavenumbers of the spawned modes are formed from
combination of any 2 wavenumbers of the first order (the same as the
toy model) and the phases follow the {\it wavenumber-phase harmonic relation}. Some terms may appear
even at the earliest stage (when $b(t) A_j A_k (\lambda_j
\lambda_k/\lambda_i)^2 > 2 A_i$). Those are terms usually involving
high-frequency modes, i.e. high $\lambda_j$ or $\lambda_k$. Such phase
coupling can manifest itself through phase mapping \cite{mapping}. For
example, a short sequence of modes is formed with  a constant
difference in wavenumber $\Delta k=\lambda_k$: 
\begin{equation}
\frac{a_j a_k}{2}\bay{\lambda_j-\lambda_k}{\alpha_j - \alpha_k},
a_j \bay{\lambda_j}{\alpha_j}, \frac{a_j a_k}{2}
\bay{\lambda_j+\lambda_k}{\alpha_j + \alpha_k}, \label{eq:modespawning}
\end{equation}
where the middle mode is taken from the $\delta^{(1)}$. Due to the
wavenumber-phase harmonic relation, this phase sequence has a constant
phase difference $\Delta \phi =\alpha_k$ and can be mapped along a line parallel to
the diagonal through phase mapping technique. Such coupling between
modes with large $\Delta{\bi k}$ is discussed in Chiang, Coles \&
Naselsky \shortcite{mapping}.  

\item {\bf Mode merging :} The newly--spawned modes from the
second order are not all independent but could merge with modes
of the same wavenumbers in the first order. Take the following two
modes from Eq.(\ref{eq:quadraticfield}) as an example, one from the
$\delta^{(1)}$ and the other from the $\delta^{(2)}$: 
\begin{equation}
 a_i\bay{\lambda_i}{\alpha_i}, \frac{a_j
 a_k}{2}\bay{\lambda_j-\lambda_k}{\alpha_j-\alpha_k}. \label{eq:modecoupling}
\end{equation} 
If $\lambda_j-\lambda_k =\lambda_i$, i.e. the wavenumber of the
newly-spawned mode coincides with the one in the first order, and when
$b(t)$ reaches a stage $b(t)\simeq 2 (A_i /A_j
A_k)(\lambda_i/\lambda_j \lambda_k)^2$  such that $a_i \simeq a_j a_k
/2$, these two modes merge to form a new mode: 
\begin{equation}
2 a_i \cos [(\alpha_j-\alpha_k
-\alpha_i)/2]\bay{\lambda_i}{\frac{\alpha_j - \alpha_k + \alpha_i}{2}},
\end{equation}  
where the amplitude is modulated and the phase is
shifted. Such modulation in amplitudes and phases proceeds
gradually and the consequence is that the wavenumber-phase harmonic
relation is broken. Thus not all spawned
modes will enjoy the harmonic relation. Note
that in this case the evolution can still be at very early stage $b(t)
\ll 1$ as long as $\lambda_j \lambda_k/\lambda_i \gg (A_i/A_jA_k)^{1/2}$.  
\end{itemize}

We can also look at the density contrast up to the third order,  
\begin{eqnarray}
\delta^{(1)} + \delta^{(2)}+ \delta^{(3)} & = & \sum_i a_i \bay{\lambda_i}{\alpha_i} + \sum_{jk}
\frac{a_j a_k}{2} \left[\bay{\lambda_j - \lambda_k}{\alpha_j -
\alpha_k} + \bay{\lambda_j + \lambda_k}{\alpha_j + \alpha_k} \right]
\nonumber \\
& + & \sum_{\ell m n}\frac{a_\ell a_m a_n}{4}  
\left[\bay{\lambda_\ell+\lambda_m+\lambda_n}{\alpha_\ell+\alpha_m+\alpha_n} + 
      \bay{\lambda_\ell+\lambda_m-\lambda_n}{\alpha_\ell+\alpha_m-\alpha_n} + 
      \bay{\lambda_\ell-\lambda_m+\lambda_n}{\alpha_\ell-\alpha_m+\alpha_n} \right. \nonumber \\
&  + & \left. \bay{\lambda_\ell-\lambda_m-\lambda_n}{\alpha_\ell-\alpha_m-\alpha_n} 
\right]. \label{eq:thirdorder}
\end{eqnarray}
Once again new modes are spawned, more complicated phase coupling
appears and the spawned modes also merge with parent modes. We shall
look at mode spawning and coupling at higher-order modes through direct
simulations.

\subsection{Bispectrum and quadratic phase coupling}
Bispectrum is the lowest-order statistic sensitive to the structure
generated by gravitational clustering \cite{peebles}. It is defined as
two-dimensional Fourier transform of the connected three-point
auto-covariance function $\zeta$
\begin{equation}
\zeta(r_1, r_2)= \langle
\delta(x)\delta(x+r_1)\delta(x+r_2)\rangle.
\end{equation}
The bispectrum and higher-order polyspectra vanish for Gaussian random
fields, but in a non-Gaussian field they may be non-zero. The
usefulness of these and related quantities therefore lies in the fact
that they encode some information about non-linearity and
non-Gaussianity. 

A certain form of phase relationship produces non-zero bispectrum. To
see this, we simplify the $\delta^{(2)}$ of Eq.(\ref{eq:2order}) by
taking only the 2nd, 3rd and 6th term and denote it as the case I \cite{proceeding}, 
\begin{equation}
\delta_{\rm I} = a_1 \bay{\lambda_1}{\alpha_1} + a_2
\bay{\lambda_2}{\alpha_2} + a_1 a_2 \bay{\lambda_1+\lambda_2}{\alpha_1
+ \alpha_2}. \label{eq:model1}
\end{equation}
For comparison we introduce the case II:
\begin{equation}
\delta_{\rm II} = a_1 \bay{\lambda_1}{\alpha_1} + a_2
\bay{\lambda_2}{\alpha_2} + a_1 a_2 \bay{\lambda_1 +
\lambda_2}{\alpha_3},
\label{eq:model2}
\end{equation}
where $\alpha_1$, $\alpha_2$ and $\alpha_3$ are random. The case I
displays {\it quadratic phase coupling} whereas the case II exhibits no phase
association. One can see that $\langle \delta_{\rm I}\rangle=\langle
\delta_{\rm II}\rangle=0$, and it is straightforward to show that the
covariances for both cases are equal, 
\begin{equation}
\xi_{\rm I}(r)  =  \langle\delta_{\rm I} (x) \delta_{\rm
I}(x+r)\rangle = \xi_{\rm II}(r)
=  \frac{a_1^2}{2} \cos (\lambda_1 r) +
\frac{a_2^2}{2}\cos (\lambda_2 r)  +  \frac{a_1^2 a_2^2}{2} \cos
[(\lambda_1+\lambda_2) r],
\end{equation}
so are their power spectra. The above demonstrates that second-order
statistics are blind to any phase associations.

The reduced three-point covariance function for the case I is 
\begin{eqnarray}
\zeta_{\rm I}(r_1,r_2) & = &  \frac{a_1^2 a_2^2}{4}
         [ \cos(\lambda_2 r_1 + \lambda_1 r_2) 
         + \cos(\lambda_1 r_1 + \lambda_2 r_1 - \lambda_1 r_2)  
         + \cos(\lambda_1 r_1 + \lambda_2 r_1 - \lambda_2 r_2) \nonumber \\ 
&&      \;\;\;\;\;\;  + \cos(\lambda_1 r_1 + \lambda_2 r_2)
         + \cos(\lambda_1 r_2 + \lambda_2 r_2 - \lambda_1 r_1)
         + \cos(\lambda_1 r_2 + \lambda_2 r_2 - \lambda_2 r_1) ], 
\end{eqnarray}
and for arbitrary $\alpha_1$, $\alpha_2$ and  $\alpha_3$ we get
\begin{equation}
\zeta_{\rm II}(r_1, r_2)=0.
\end{equation}
One can see that the spawned modes of the second order which benefit from the
wavenumber-phase harmonic
relation induce non-zero $\zeta$.
The bispectrum, $B(k_1,k_2)$, the two-dimensional
Fourier transform of $\zeta$, for case II is $B_{\rm II}(k_1,k_2)=0$ trivially,
whereas $B_{\rm I}(k_1, k_2)$ consists of a single spike located
somewhere in the region of $(k_1, k_2)$ space defined by $k_2\geq
0$, $k_1\geq k_2$ and $k_1+k_2\leq \pi$. If $\lambda_1\geq
\lambda_2$ then the spike appears at $k_1=\lambda_1$,
$k_2=\lambda_2$. Thus the bispectrum is the lowest order of
polyspectra that measures the phase coupling induced by quadratic
non-linearities when the wavenumber-phase harmonic relation
holds. However, bispectrum is trivially zero again if such harmonic relation is
broken. Note that this relation is easily broken by mode merging when
a third phase is involved.

Following the same line of thought, we can extend this toy model from
Eq.(\ref{eq:thirdorder}) to the case III,
\begin{equation}
\delta_{\rm III} = a_1 \bay{\lambda_1}{\alpha_1} + a_2
\bay{\lambda_2}{\alpha_2} +  a_3 \bay{\lambda_3}{\alpha_3} + \frac{a_1
a_2 a_3}{4} \bay{\lambda_1+\lambda_2 +\lambda_3}{\alpha_1 + \alpha_2 +
\alpha_3}, \label{eq:model3}
\end{equation}
and the case IV for comparison,
\begin{equation}
\delta_{\rm IV} = a_1 \bay{\lambda_1}{\alpha_1} + a_2
\bay{\lambda_2}{\alpha_2} + a_3 \bay{\lambda_3}{\alpha_3} + \frac{a_1
a_2 a_3}{4} \bay{\lambda_1+\lambda_2 +\lambda_3}{\alpha_4} ,
\label{eq:model4}
\end{equation}
where $\alpha_1$, $\alpha_2$, $\alpha_3$ and $\alpha_4$ are
random. The bispectrum for both cases are zero, but the trispectrum,
the three-dimensional Fourier transform of the reduced four-point
covariance function, can pick up cubic phase coupling for
case III (24 terms), whereas it is again trivially zero for case IV. The third-order term produces cubic phase coupling to which bispectrum is
blind.

\begin{figure}
\epsfig{file=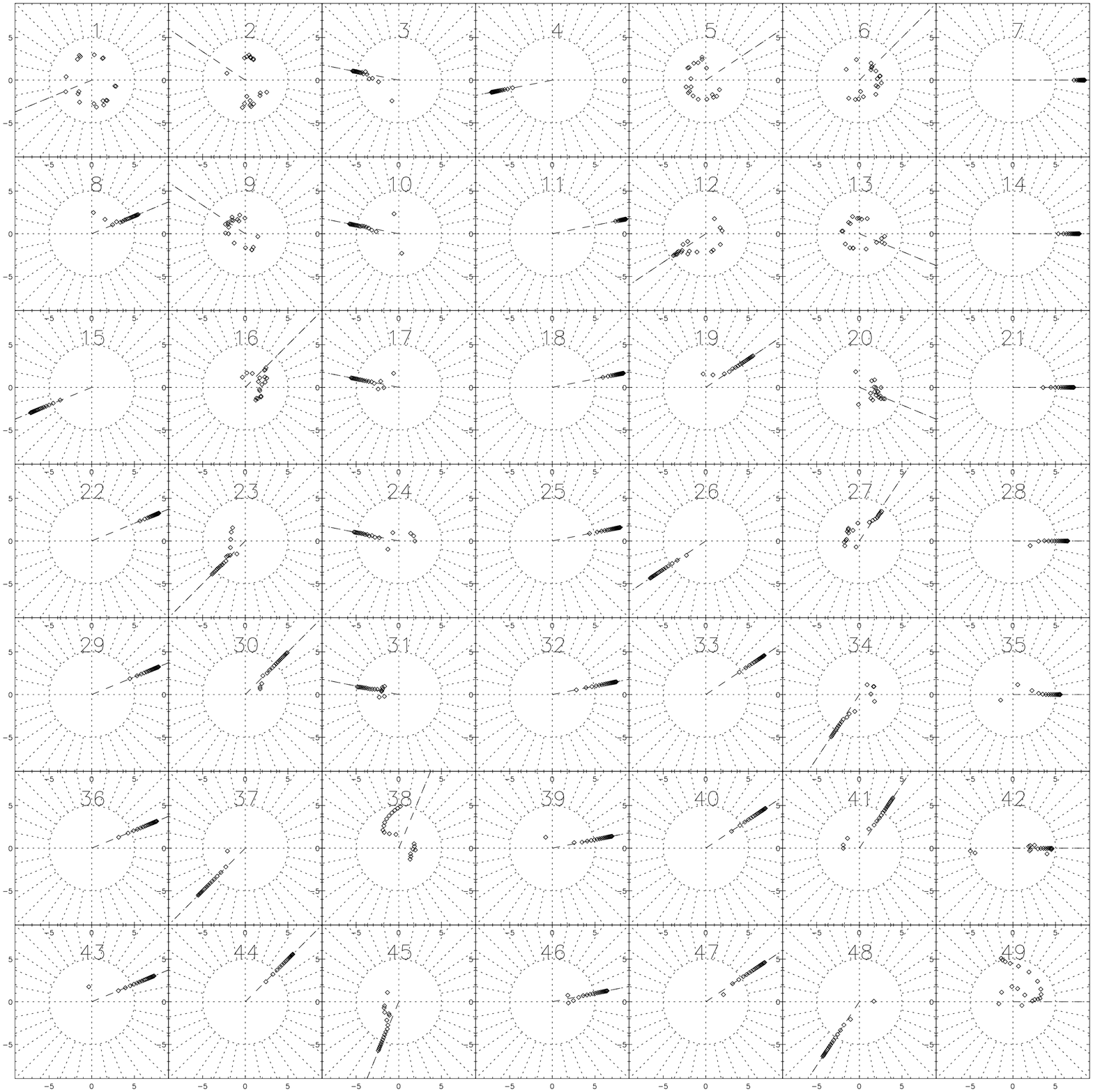,width=18cm}
\caption{The amplitude and phase evolution of
1D ZA simulation with $\Phi_0= -\cos(7 q ) - \cos(11 q +
\pi/16)$ for wavenumber $k=1-49$. We specifically designate 7 small
panels in one row so that in each column the increment of the
wavenumber is 7. Each time step is represented by a
diamond symbol. In each panel (for one
Fourier mode) the amplitude ($\times
10^{9}$) is shown in logarithmic scale by the distance to the origin, and the phase by the angle against the positive
$x$ axis. Thus the dotted circle with radius 5 units indicates the level of
amplitudes with $10^{-4}$. The radial dotted lines are angles with
increment of $\pi/16$ to show that phases align with angles of
multiple of $\pi/16$. The dash lines are the predicted phases according
to Table 1: the wavenumber-phase harmonic relation.  One can see the
phases of spawned modes follow this 
harmonic relation because they have only one primary combination
from parent modes, i.e. non-interactive.} \label{7_11.1}
\end{figure}

Therefore, bispectrum only measures phase coupling at the
second-order. During density clustering, however, mode spawning will
produce the bispectrum from case I, $\zeta_{\rm I}(r_1,r_2)$ is not
always non-zero. Mode merging breaking the wavenumber-phase harmonic relation
results in zero bispectrum. The same would happen to all
polyspectra.

Although we only demonstrate with 1D toy model the relationship between
bispectrum and quadratic phase coupling, a full treatment of such
relationship is elaborated in Watts \& Coles \shortcite{wc}: a
non-zero bispectrum can only be produced when the ${\bi k}$ vectors
form a triangle in $k$-space, wherein the same (quadratic)
wavenumber-phase harmonic relation has to hold. Phase coupling is not
registered only in bispectrum, but in all polyspectra.

\section{Simulations} \label{simulation} 
In this Section, we perform 1D ZA numerically in order to demonstrate
mode spawning and coupling into highly non-linear regime. Due to
limitation on integration in Lagrangian coordinate, in the previous
Section we can 
analyze only the onset of non-linearity by assuming $b(t) \ll
1$. In the following simulations we assume the periodic boundary condition,
\begin{equation}
x(q+ 2 j \pi)=x(q).
\end{equation}
\subsection{Non-interactive and interactive parent modes in the simulations}
Here we supply three different initial velocity
potential functions to the simulations. We choose the velocity potential field
with the following form for Fig.~\ref{7_11.1} and \ref{7_11.2}:
\begin{equation}
\Phi_0= - \cos(7 q + \theta)  -  \cos(11 q + \phi), \label{eq:noninteractive}
\end{equation}  
where $\theta$ and $\phi$ are 0 and $\pi/16$ in Fig.~\ref{7_11.1} and
$\pi/4$ and $\pi/7$ in Fig.~\ref{7_11.2}, respectively. As described
in previous Sections, once the clustering proceeds, $\theta$ and
$\phi$ will feed as phases of two parent modes with wavenumber 7 and
11 in these cases: $\alpha_7$ and $\alpha_{11}$. These specially chosen wavenumbers $\lambda_1=7$ and $\lambda_2=11$ for both
Fig.~\ref{7_11.1} and \ref{7_11.2} are ``non-interactive'' in the
sense that the phases of spawned modes have only one primary combination from
the parent ones. If the parent modes are ``non-interactive'',
the phases will eventually follow the harmonic relationship (see
Table~\ref{harmonic}). So the phase of the spawned mode with, for example, 
wavenumber $k=25$ ( $= 2 \times 7 + 11$ ) can only come from addition of the
phases of the parent modes, in other words, $\alpha_{25}=2
\times \alpha_{7} + \alpha_{11}$. 

In addition to the two simulations of
``non-interactive'' mode spawning, in Fig.~\ref{7_11_14.1} we also
show the simulation with one extra ``interactive'' mode: 
\begin{equation}
\Phi_0= - \cos(7 q )  -  \cos(11 q + \pi/16) - \cos(14 q + \pi/4). \label{eq:interactive}
\end{equation}  
Figure~\ref{7_11_14.1} with the extra mode  $\lambda_3 =14 =
2\lambda_1$ illustrates parent-mode interactions leading to change of
phases during the clustering process. One can see now, with this extra
parent mode, the spawned wavenumber $k = 25$, for example, can have two combinations: $2 \times 7 +11$ and $
11 +14$. So the phase of this spawned mode $\alpha_{25}$ can now come
from either $2 \alpha_{7} +\alpha_{11}$ or $\alpha_{11} + \alpha_{14}$.

We using the following representation in order to display
simultaneously the amplitude and phase evolution for each Fourier
mode. Each small panel
represents one Fourier mode with denoted wavenumber. We divide the
simulations (up to the shell crossing) into 20 time steps with each step
represented by one diamond symbol. In each panel the amplitudes of different
time steps are shown in logarithmic scale by the distances to the
origin, and the phases by the angles against the positive $x$ axis
(similar to a complex-plane representation but only the magnitudes in
logarithmic scale). We have to multiply all the amplitudes by $10^{9}$ so
that they are always positive after taking the
logarithm\footnote{This is to avoid the degeneracy between
negative amplitudes (after taking the logarithm) with positive phases
and positive amplitudes with opposite phases.}. Thus the
dotted circles with radius 5 units in each panel 
indicate a level of Fourier amplitudes $10^{-4}$. 

In Fig.~\ref{7_11.1} the radial dotted lines are angles with increments
of $\pi/16$ to show that phases aligned with angles of multiple
$\pi/16$ (which is due to combination of $\alpha_7=0$ and
$\alpha_{11}=\pi/16$). The dash lines are the predicted phases from
wavenumber-phase harmonic relation according to Table~\ref{harmonic}.

One can easily see in Fig.\ref{7_11.1} that the parent modes $k=7$
and 11 have higher initial amplitudes than other modes for the
start. We specifically designate 7 small panels in one row so that in each
column the increment of the wavenumber is 7. Because wavenumber 7 is
one of the parent modes, phases of the spawned modes are formed by
addition of $\alpha_7$, following wavenumber-phase harmonic
relation. For instance, for the wavenumbers of the Fourier modes
18 ($ = 11 + 7$), 25 ($=11 + 2 \times 7 $), 32 ($=11 + 3 \times 7 $), 39
($=11 + 4 \times 7 $), 46 ($=11 + 5 \times 7 $), their phases are formed from
$\alpha_{11}+\alpha_7$, $\alpha_{11}+ 2 \times \alpha_7$,
$\alpha_{11}+ 3 \times \alpha_7$, $\alpha_{11}+ 4 \times \alpha_7$ and
$\alpha_{11}+ 5 \times \alpha_7$, an increment of $\alpha_7$ (see
Table~\ref{harmonic} for the details of phase combination from the
parent modes). It is therefore easy to see that the multiples of
wavenumber 7: 7, 14, 28, 35, 42  have the same phase, as well as 8, 15, 22, 29,
36, 43. This is due to the fact that the phases formed by the
increment of $\alpha_7$ is unchanged when $\alpha_7=0$ in Fig.\ref{7_11.1}. Phases of most
modes eventually align with multiples of $\pi/16$ because of the
following reasons: they follow the wavenumber-phase harmonic relation
and the phase of one parent mode is 0. Modes with wavenumbers $k=1$,
2, 5, 6, 9 and 13 do not align with multiples of $\pi/16$ and their
amplitudes cannot reach $10^{-4}$.

\subsection{Analytical account for the simulations} 
To account for the interaction of the parent modes leading to mode
merging, we can start from Eq.~(\ref{eq:delta1d}). We have 
\begin{equation}
\delta_k=\frac{1}{2\pi}\int^{\pi}_{-\pi}
\left[\left(\frac{dx}{dq}\right)^{-1} - 1\right]e^{-\imath k x}
dx= \frac{1}{2\pi}\int^{\pi}_{-\pi}e^{-\imath kx} dq =
\frac{1}{2\pi}\int^{\pi}_{-\pi}\exp \{- \imath k [q+b(t)
u(q) ] \} dq, \label{eq:integral}
\end{equation}
where $u(q)=-d\Phi_0(q)/d q$. We can give the same treatment to
Eq.~(\ref{eq:integral}) as Eq.~(\ref{eq:expansion}) by expressing it in
terms of a power series of $b(t)$: 
\begin{equation}
\delta_k=\sum^{\infty}_{p=0} \frac{(-\imath k b)^p}{p \; !}\frac{1}{2\pi}
\int^{\pi}_{-\pi} dq \; e^{-\imath kq} u(q)^p.
\end{equation}
Following Soda \& Suto \shortcite{sodasuto}, we can denote the Fourier
transform of $u(q)$ as $u_k$
\begin{equation}
\uk=\frac{1}{2\pi}\int^{\pi}_{-\pi} dq\; u(q) e^{-\imath k q},
\end{equation}
so the Fourier modes $\delta_k$ can be expressed as 
\begin{equation}
\delta_k= -\imath kb \uk + \frac{(-\imath kb)^2}{2!}\sum_{k_1,k_2} u_{k_1} u_{k_2}
\delta^D[k-(k_1+k_2)] + \frac{(-\imath kb)^3}{3!}\sum_{k_1,k_2,k_3} u_{k_1}
u_{k_2}u_{k_3} \delta^D[k-(k_1+k_2+k_3)]+\cdots, \label{eq:series}
\end{equation}
where $\delta^D$ denotes Dirac-delta function. For 
$u(q)=-d\Phi_0/dq=- \sum_j A_j \lambda_j \sin(\lambda_j q +\alpha_j)$
from the velocity potential $\Phi_0=- \sum_j A_j \cos( \lambda_j q +\alpha_j)$,  
\begin{equation}
u_k = -\frac{1}{2\imath}\sum_j A_j \lambda_j e^{\imath \alpha_j}
\delta^D(\lambda_j -k)=-\frac{A_k \lambda_k}{2\imath} e^{\imath
\alpha_k}=\frac{A_k \lambda_k}{2} e^{\imath(\pi/2 + \alpha_k)}, \label{eq:uk}
\end{equation}
and $u_{-k}$ is simply its complex conjugate. 

Equation~(\ref{eq:series}) and (\ref{eq:uk}) provide us the insight into
mode spawning and mode merging in
gravitational clustering. The Dirac-delta functions dictate the
spawning of Fourier modes, both the amplitudes and phases. If, for any
Fourier mode $k$, there is only
one combination from the parent modes, its phase $\alpha_k$ shall not
change and the amplitude grow steadily with $b(t)^{n}$ (according to
the order of this combination). If, on the other hand, there is more than one set of combination for wavenumber $k$ [cf
Eq.~(\ref{eq:modecoupling})], they will interact so that the phase
$\alpha_k$ will be changed accordingly.

\begin{table}
\centering
\begin{tabular}{ccccccc}
\hline $\alpha_1=$ & $\alpha_2=$ & $\alpha_3=$ & $\alpha_4=$ & $\alpha_5=$
& $\alpha_{6}=$ & $\alpha_7$ \\
\fbox{$-3\alpha_7+2\alpha_{11}+\pi$}  &
\fbox{$5\alpha_7-3\alpha_{11}+\pi$} & $2\alpha_7-\alpha_{11}+\pi$ &
$-\alpha_7+\alpha_{11}+\pi$ & \fbox{$-4\alpha_7+ 3\alpha_{11}$} &
\fbox{$- 4 \alpha_{7} +2 \alpha_{11}$} &  \\
\hline $\alpha_8=$ & $\alpha_9=$ & $\alpha_{10}=$ & $\alpha_{11}$ &
$\alpha_{12}=$ &  $\alpha_{13}=$ & $\alpha_{14}=$ \\
$-2\alpha_{7}+2\alpha_{11}$ & \fbox{$6\alpha_7-3\alpha_{11}+\pi$}
&$3 \alpha_7-\alpha_{11}$ &  &
$-3\alpha_{7}+3\alpha_{11}+\pi$ & \fbox{$5\alpha_{7} - 2\alpha_{11}$}
 &  $2\alpha_{7}$ \\
\hline $\alpha_{15}=$ &  $\alpha_{16}=$ & $\alpha_{17}=$ & $\alpha_{18}=$
& $\alpha_{19}=$ & $\alpha_{20}=$ & $\alpha_{21}=$ \\
$-\alpha_{7}+ 2\alpha_{11}+\pi$ & $-
4 \alpha_{7} +4 \alpha_{11}$ & $4\alpha_{7} -\alpha_{11}+\pi$ & $\alpha_{7}
+\alpha_{11}$ & $-2\alpha_{7} +3 \alpha_{11}$ & $6\alpha_{7} -2
\alpha_{11}$  & $3\alpha_{7}$ \\
\hline $\alpha_{22}=$ & $
\alpha_{23}=$ & $\alpha_{24}=$ & $\alpha_{25}=$ & $\alpha_{26}=$ & $ \alpha_{27}=$ & $ \alpha_{28}=$ \\ 
$2 \alpha_{11}$ & $-3\alpha_{7} +4 \alpha_{11}+\pi$ & $5\alpha_{7} -
\alpha_{11}+\pi$ & $2\alpha_{7} + \alpha_{11}$ & $-\alpha_{7} +3 \alpha_{11}+\pi$ & $-4\alpha_{7} +5 \alpha_{11}$ & $4\alpha_{7}$ \\ 
\hline $\alpha_{29}=$ &  $\alpha_{30}=$ & $\alpha_{31}=$ &  $\alpha_{32}=$
&  $\alpha_{33}=$ & $\alpha_{34}=$ &  $\alpha_{35}=$ \\
$\alpha_{7} +2 \alpha_{11}$ &  $-2\alpha_{7} +4 \alpha_{11}$ & $6\alpha_{7} - \alpha_{11}+\pi$ &  $3\alpha_{7} + \alpha_{11}$ &  $3 \alpha_{11}$ &
$-3 \alpha_{7} +5 \alpha_{11}+\pi$ &  $5 \alpha_{7}$ \\
\hline  $\alpha_{36}=$ &  $\alpha_{37}=$ & $\alpha_{38}=$ &  $\alpha_{39}=$ &  $\alpha_{40}=$ &  $\alpha_{41}=$ &  $\alpha_{42}=$ \\
 $2 \alpha_{7} +2 \alpha_{11}$ &  $-
\alpha_{7} +4 \alpha_{11}+\pi$ & \fbox{$-
4 \alpha_{7} +6 \alpha_{11}$} &  $4 \alpha_{7} +
\alpha_{11}$ &  $ \alpha_{7} +3 \alpha_{11}$ &  $-2 \alpha_{7} +5
 \alpha_{11}$ &  $6 \alpha_{7}$ \\
\hline  $\alpha_{43}=$ & $\alpha_{44}=$ &  $\alpha_{45}=$ & $\alpha_{46}=$ &
$\alpha_{47}=$ &   $\alpha_{48}=$ & $\alpha_{49}=$ \\
$3 \alpha_{7} +2 \alpha_{11}$ &
$4 \alpha_{11}$ &  $-3 \alpha_{7} +6
\alpha_{11}+\pi$ & $5 \alpha_{7} + \alpha_{11}$ &
$2 \alpha_{7} + 3\alpha_{11}$ &   $-\alpha_{7} +5 \alpha_{11}+\pi$ & \fbox{$ 7 \alpha_{7}$} \\
\hline
\end{tabular}
\caption{The predicted phases from the harmonic relation for parent
wavenumber 7 and 11 for Fig.\ref{7_11.1} and \ref{7_11.2}. The terms
with odd number of complex conjugate $u_k$ give rise to an extra phase
shift by $\pi$, apart from the harmonic relation. The spawned modes that have
combinations from the parent wavenumbers (with lower orders) will grow,
with phases following the same harmonic relation. Those suppressed are
modes which need high orders of combinations from
wavenumbers of the parent modes (those in frames, e.g. $k=13$ and 38).}
\label{harmonic} 
\end{table}

To be more specific, in our case of two cosine functions in
$\Phi_0(q)$ (in Fig.1 and 2) from which only $u_{\lambda_a}$ and
$u_{\lambda_b}$ exist ($\lambda_b >\lambda_a$), we can write down
explicitly what constitutes the Fourier mode of wavenumber
$k$: the first term in Eq.~(\ref{eq:series}) is simply $A_k
\lambda_k k b e^{\imath \alpha_k}/2$. The second term has all the
combinations of any two from $\lambda_a$ and $\lambda_b$ in the
Dirac-delta functions: $2 \lambda_a$, $2 \lambda_b$, $\lambda_a
+\lambda_b$ and $\lambda_a - \lambda_b$ so that
\begin{eqnarray}
&& \frac{(-\imath kb)^2}{2!}\sum_{k_1,k_2} u_{k_1} u_{k_2}
\delta^D[k-(k_1+k_2)]   \nonumber \\
& = & \frac{k^2 b^2 e^{\imath \pi}}{2} \left[ \frac{A_{\lambda_a} A_{\lambda_b}
\lambda_{a}\lambda_{b}}{4} e^{\imath
(\alpha_{\lambda_a}+\alpha_{\lambda_b}+\pi)} \delta^D(k-\lambda_a
-\lambda_b) + \frac{A_{\lambda_a}
A_{\lambda_b} \lambda_{a}\lambda_{b}}{4} e^{\imath
(\alpha_{\lambda_b}-\alpha_{\lambda_b})} \delta^D(k+\lambda_a -
\lambda_b) \right.\nonumber \\ 
& + & \left. \frac{A_{\lambda_a}^2 \lambda_{a}^2}{4} e^{\imath (2
\alpha_{\lambda_a}+\pi)} \delta^D(k-2 \lambda_a)+ \frac{A_{\lambda_b}^2
\lambda_{b}^2}{4} e^{\imath (2 \alpha_{\lambda_b}+\pi)} \delta^D(k-2
\lambda_b)\right]. \label{eq:2ndterm}
\end{eqnarray}
If the $\Phi_0$ has more than two cosine functions, the second term of 
Eq.~(\ref{eq:series}) then shall have all the combinations of any 2
from $\Phi_0$. The third term involves combinations of any 3 from
$\lambda_a$ and $\lambda_b$, i.e. $3\lambda_a$, $3\lambda_b$,
$2\lambda_a \pm \lambda_b$, $\lambda_a \pm 2\lambda_b$. It is also
worth noting that unless the higher-order terms
exist (i.e. $\delta^D[k-(k_1+k_2+k_3)]=1$), the phases of the Fourier
modes are fixed and the amplitudes grow with $b^{n}$.

For the interactive parent modes in Eq.(\ref{eq:interactive})
(Fig.~\ref{7_11_14.1}), where $\lambda_3 = 2 \lambda_1$,
the second term of Eq.(\ref{eq:series}) has one Dirac-delta function involving
$\lambda_3+\lambda_2$ whereas in the third term there is another involving
$2\lambda_1 +\lambda_2$. The resultant phase of this wavenumber
depends on the amplitudes of the terms involving the two Dirac-delta
functions.

Note that for wavenumber $k=\lambda_b-\lambda_a$ of
Eq.(\ref{eq:2ndterm}) the phase is
$\alpha_{\lambda_b}-\alpha_{\lambda_a}+\pi$. This extra phase shift by
$\pi$ is due to odd number of complex conjugate of $u_k$ in
Eq.~(\ref{eq:series}).
\begin{figure}
\epsfig{file=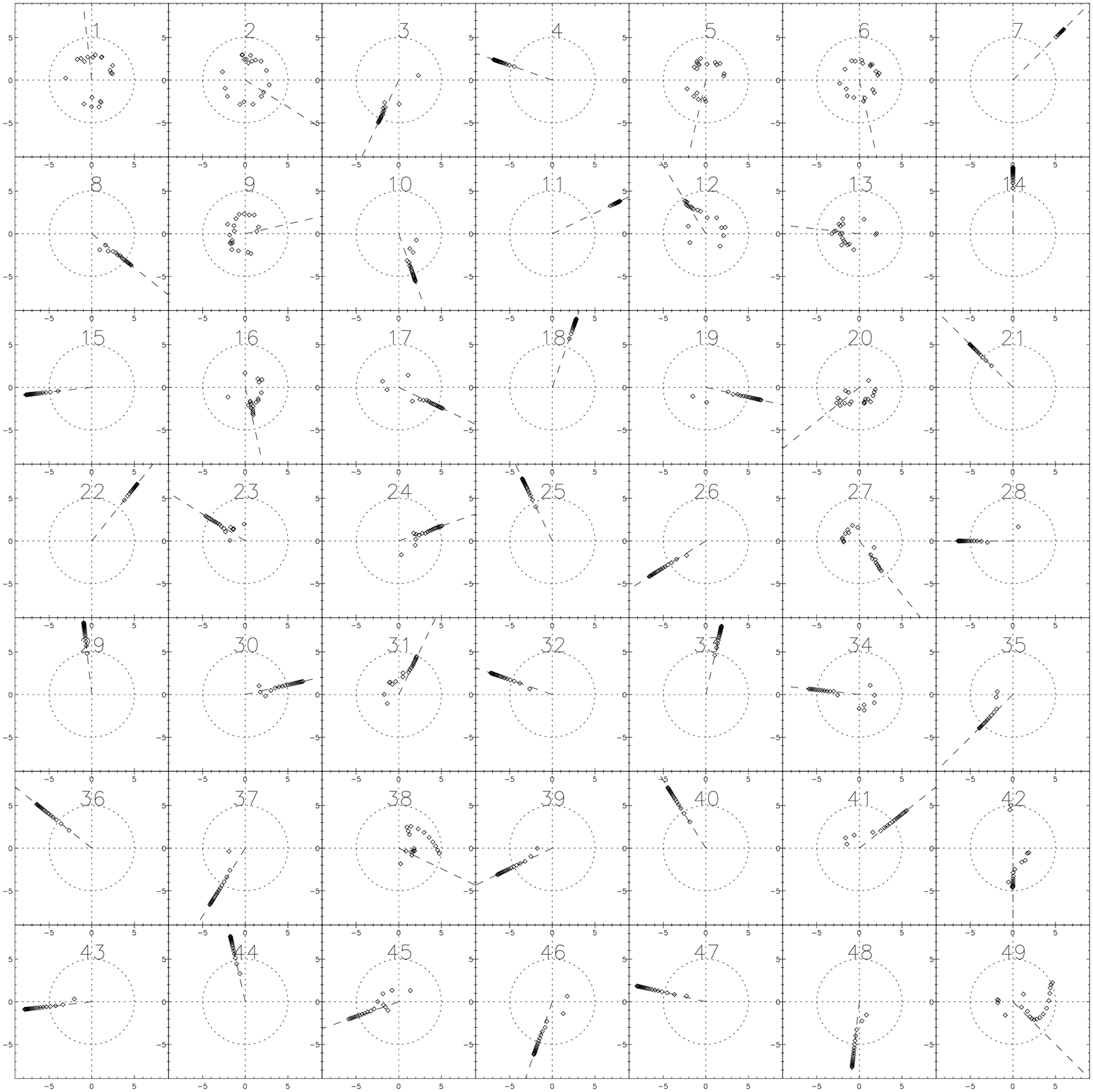,width=18cm}
\caption{The amplitude and phase evolution (wavenumber $k=1-49$) of
1D ZA simulation with $\Phi_0= -\cos(7 q +\pi/4) - \cos(11 q +
\pi/7)$. In each panel the amplitude ($\times
10^{9}$) is shown in logarithmic scale by the distance of each
symbol to the origin, and the phase by the angle against the positive
$x$ axis. Thus the dotted circle in each panel indicates amplitude with
$10^{-4}$. The dash lines indicate the predicted phases according to
Table 1: the wavenumber-phase harmonic relation.} \label{7_11.2}
\end{figure}

\begin{figure}
\epsfig{file=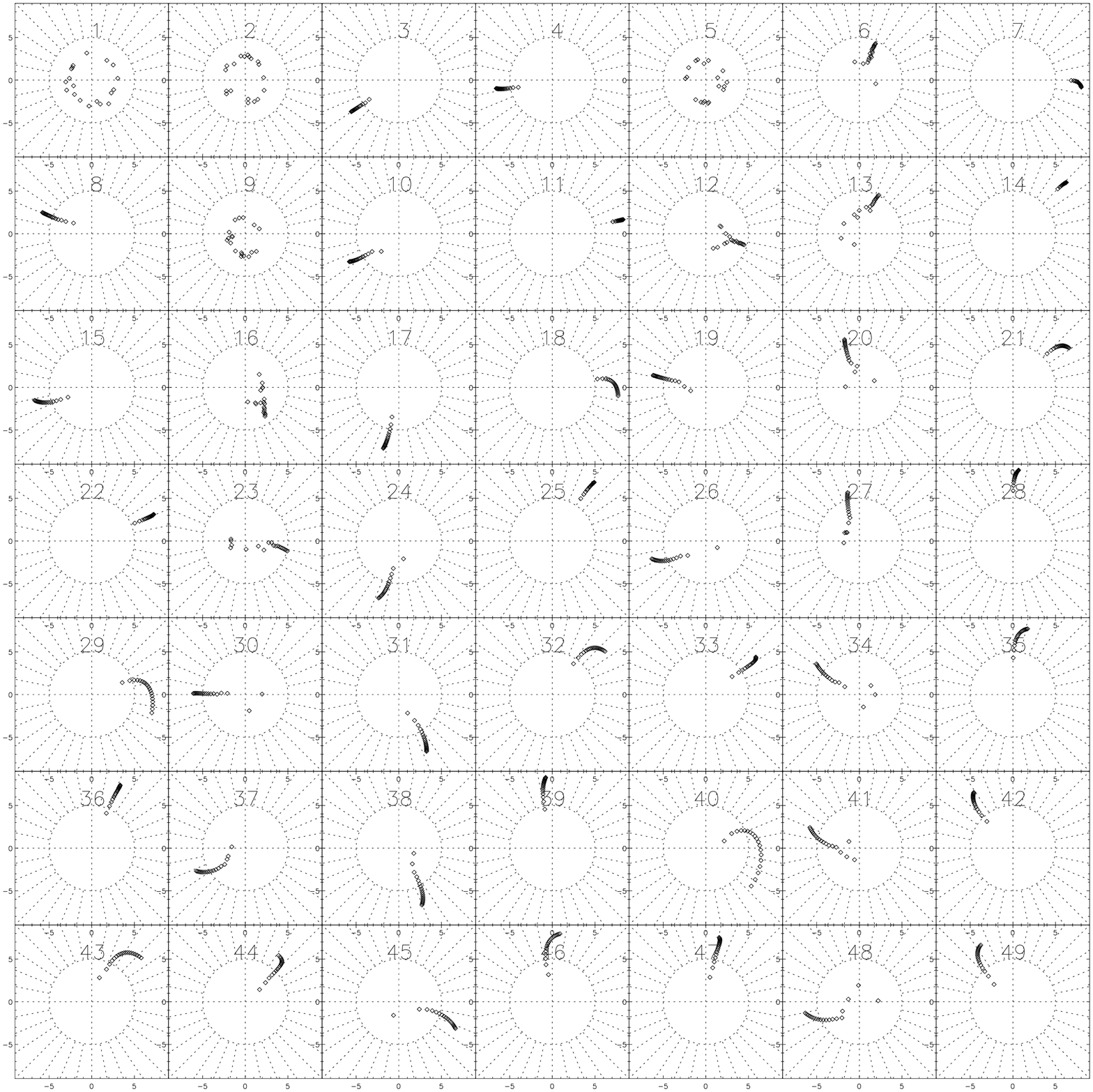,width=18cm}
\caption{The amplitude and phase evolution of
1D ZA simulation with $\Phi_0=- \cos(7 q) - \cos(11 q +\pi/16) - \cos(14 q +
\pi/4)$. As in Fig.~\ref{7_11.1} the radial dotted lines are angles
with an increment of $\pi/16$ to show that phases align with angles of
multiple of $\pi/16$. One can see now, with the extra parent mode
$\lambda_3=14$ interacting, the phases are shifted considerably. Note
also that in this case, the wavenumber-phase harmonic relation is broken.} 
\label{7_11_14.1}
\end{figure} 

In Table~\ref{harmonic} we show the predicted phases of the spawned
Fourier modes from the parent modes for
Fig.~\ref{7_11.1} and \ref{7_11.2}. The wavenumbers of the parent
modes are 7 and 11. The spawned modes that have the combination of the
parent wavenumber (with lower orders) will grow following the
wavenumber-phase harmonic relation. Those which are suppressed (or have not grown before shell crossing) are modes which need high orders of
combinations in wavenumbers or which have no exact combination. The
phase of wavenumber 24: $\alpha_{24}$, for example, can
have combinations of $5\alpha_{7}-\alpha_{11}$ or $-6 \alpha_7 + 6
\alpha_{11}$. The former is of the 6th order and the latter is the
12th order in terms of $b(t)$ series in Eq.(\ref{eq:series}). The
low-order combination of wavenumbers prevails at the early stage and
if it is dominant in amplitude, the other combinations will have little
influence. When the parent modes are ``interactive'', the phase sets
out with a value from lower-order combination. In Fig.~\ref{7_11_14.1}
we have 3 parent modes $\lambda_1=7$, $\lambda_2=11$ and
$\lambda_3=14$. The phase of mode $k=18$ is spawned from $\alpha_7 +
\alpha_{11}$ in the beginning and
$3\alpha_{7}+\alpha_{11}-\alpha_{14}$ at later stage. It is only at
very later stage of evolution that the 5th order can catch up with the
2nd order (see Fig.~\ref{7_11_14.1}).  

Having such rules in mind, one can clearly see why the parent low-frequency
modes (and their phases) have decisive influence on the mode
coupling. As shown in Chiang \& Coles \shortcite{phaseentropy}, convergence of
phase difference $D_k=\phi_{k+1}-\phi_k$ at high $k$ is related to the
location of the highest density peak. The wavenumbers of parent
lowest-frequency mode ($k=1$) can easily be added (or subtracted) to
form precisely the spawned wavenumbers, particularly for high-frequency
modes, so the phase of the lowest $k$ mode would inevitably take part in
most of the hight $k$ mode spawning. Hence the increment in phases
(i.e. $D_k$) along $k$ axis would be the phase from the lowest frequency
mode ($k=1$).  

Note also that for more generic situations where all
modes in $\Phi_0$ have non-zero amplitudes, they are definitely
``interactive'', in particular with the presence of
fundamental mode $k=1$. Therefore, the evolution of amplitudes and
phases shall have a complicated picture with mode merging with
changing of Fourier amplitudes and phases.

\section{Discussions} 
In this paper, we use 1D ZA to illustrate some non-linear
effects induced by gravitational clustering: mode spawning
and mode coupling. We first demonstrate the onset of
non-linearity by expanding the density contrast in terms of power series
of $b(t)$. The second-order of such expansion reflects quadratic
density fields, from which modes are spawned with phases following a
special wavenumber-phase harmonic relation. We have also illustrated with toy
models that the widely-used statistic, bispectrum, can only pick up
the quadratic phase coupling and is blind to cubic and higher-order phase
coupling. We further perform direct 1D ZA simulations in order to
demonstrate these effects entering non-linear regime. The
wavenumber-phase harmonic relation holds as long as the wavenumbers of
parent modes are not ``interactive'', as shown in Fig.~\ref{7_11.1} and
\ref{7_11.2}. The complexity of Fourier amplitude and phase evolution
comes from the continuous interactions between modes and their
merging and coupling with the spawned modes.  

Recently it is also claimed that bispectrum measures phase
correlations \cite{wmapng}, which is partially correct. We have
demonstrated with toy models that bispectrum measures only {\it
quadratic} phase coupling under the condition that the
wavenumber-phase harmonic relation holds, as pointed out in details in Watts \&
Coles \shortcite{wc}. Upon using bispectrum as a test of
non-Gaussianity, a quadratic field such as
the form $\delta({\bi x})+ \epsilon \delta^2({\bi x})$ could give a
zero bispectrum due to the interaction between the spawned modes in
$\epsilon \delta^2({\bi x})$ with the parent ones in $\delta({\bi
x})$, hence resulting in a false signature of Gaussianity. Moreover,
the blindness of bispectrum to higher-order phase coupling indicates
that a complete hierarchy of polyspectra is required to fully
characterize the statistical properties of a fluctuation field.

The initial power spectral index decides the clustering
morphology, which is closely related to Fourier phases \cite{c3}.
It is therefore interesting to examine the link between Gaussian random fields
with different initial power spectral index, the only available
information, and evolved morphology with information encoded in
phases rather than in amplitudes. The 1D ZA simulations in this paper
can be considered as from initial white noise power spectrum with
spectral index $n=0$ (as the coefficients of cosine functions in the
velocity potential are all unity). It is clear that mode spawning is
independent of amplitudes of the parent modes, but mode merging and
coupling is rather dependent on the amplitudes of spawned modes, which
are originated from parent ones. How the information is transferred
from Fourier amplitudes to different phase coupling configurations
will be examined in the next paper.
    
\section*{Acknowledgments}
This paper was supported in part by Danmarks Grundforskningsfond
through its support for the establishment of the Theoretical 
Astrophysics Center. The author thanks Peter Coles
for useful suggestions and Pavel Naselsky for discussions.
%
%
\newcommand{\autetal}[2]{{#1\ #2. et al.,}}
\newcommand{\aut}[2]{{#1\ #2.,}}
\newcommand{\saut}[2]{{#1\ #2.,}}
\newcommand{\laut}[2]{{#1\ #2.,}}

%
%
\newcommand{\refs}[6]{#5, #2, #3, #4} 
\newcommand{\unrefs}[6]{#5, #2 #3 #4 (#6)}  

%
%

\newcommand{\book}[6]{#5, #1, #2} 
%

\newcommand{\proceeding}[6]{#5, in #3, #4, #2} 

\newcommand{\combib}[3]{\bibitem[\protect\citename{#1 }#2]{#3}} 

%
%

%
%
%
%
%

%
%
\def\apj{ApJ}
\def\apjl{ApJL}
\def\mn{MNRAS}  
\def\nature{nat} 
\def\aa{A\&A}   
\def\prl{Phys.\ Rev.\ Lett.}
\def\pr{Phys.\ Rep.}

\newcommand{\amp}{\& }
\def\cqg{Class.\ Quant.\ Grav.}
\def\grg{Gen.\ Rel.\ Grav.}

\def\cambridgepress{Cambridge University Press, Cambridge, UK} 
\def\princetonpress{Princeton University Press}
\def\worldpress{World Scientific, Singapore}
\def\oxfordpress{Oxford University Press}

\end{document}